\documentclass[aps,prl,groupedaddress,floatfix]{revtex4-1}
 \usepackage{graphicx,color,caption}

\begin{document}
\makeatletter       
  \renewcommand{\@biblabel}[1]{#1.}
  \makeatother


\title{Symbiotic Cell Differentiation and Cooperative Growth in Multicellular Aggregates}


\author{Jumpei Yamagishi}
\email[]{9464697641@mail.ecc.u-tokyo.ac.jp}
\affiliation{College of Arts and Sciences, The University of Tokyo, 3-8-1 Komaba, Meguro-ku, Tokyo 153-8902, Japan}
\author{Nen Saito}
\email[]{saito@complex.c.u-tokyo.ac.jp}
\author{Kunihiko Kaneko}
\email[]{kaneko@complex.c.u-tokyo.ac.jp}
\affiliation{Graduate School of Arts and Sciences, The University of Tokyo, 3-8-1 Komaba, Meguro-ku, Tokyo 153-8902, Japan}




\maketitle


\section{
{\large Abstract}
}
As cells grow and divide under a given environment, they become crowded and resources are limited, as seen in bacterial biofilms and multicellular aggregates. 
These cells often show strong interactions through exchanging chemicals, as evident in quorum sensing, to achieve mutualism and division of labor. 
Here, to achieve stable division of labor, three characteristics are required.
First, isogenous cells differentiate into several types. Second, this aggregate of distinct cell types shows better growth than that of isolated cells without interaction and differentiation, by achieving division of labor. 
Third, this cell aggregate is robust 
with respect to 
the number distribution of differentiated cell types. 
Indeed, theoretical studies have thus far considered how such cooperation is achieved when the ability of cell differentiation is presumed. Here, we address how cells acquire the ability of cell differentiation and division of labor simultaneously, which is also connected with the robustness of a cell society. 
For this purpose, we developed a dynamical-systems model of cells consisting of chemical components with intracellular 
catalytic 
reaction dynamics. The reactions convert external nutrients into internal components for cellular growth, and the divided cells interact through chemical diffusion. We found that cells sharing an identical catalytic network spontaneously differentiate via induction from cell-cell interactions, and then achieve division of labor, enabling a higher growth rate than that in the unicellular case. This symbiotic differentiation emerged for a class of reaction networks under the condition of nutrient limitation and strong cell-cell interactions. Then, robustness in the cell type distribution 
was 
achieved, while instability of collective growth could emerge even among the cooperative cells when the internal reserves of products 
were 
dominant. 
The present mechanism is 
simple and general as a natural consequence 
of interacting cells with limited resources, and is consistent with the observed behaviors and forms of several aggregates of unicellular organisms.
\section{
{\large Author Summary}
}
Unicellular organisms, when aggregated under limited resources, often exhibit behaviors akin to multicellular organisms, possibly without advanced regulation mechanisms, as observed in biofilms and bacterial colonies. Cells in an aggregate have to differentiate into several types that are specialized for different tasks, so that the growth rate should be enhanced by the division of labor among these cell types. To consider how a cell aggregate can acquire these properties, most theoretical studies have thus far assumed the fitness of an aggregate of cells and the ability of cell differentiation \textit{a priori}. In contrast, we developed a dynamical-systems model consisting of cells without assuming predefined fitness. The model consists of catalytic-reaction networks for cellular growth. By extensive simulations and theoretical analysis of the model, we showed that cells growing under the condition of nutrient limitation and strong cell-cell interactions can differentiate with distinct chemical compositions. They achieve cooperative division of labor by exchanging the produced chemicals to attain a higher growth rate. The conditions for spontaneous cell differentiation and collective growth of cells are presented. The uncovered symbiotic differentiation and collective growth are akin to economic theory on division of labor and comparative advantage. 
\section{
{\large Introduction}
}
As unicellular organisms grow and divide, they often form a crowded aggregate.
As exemplified by bacterial biofilms \cite{Shapiro2,MCO,Jefferson} 
and slime molds \cite{CSM, CSM2}, these aggregates are not merely crowded passively, but sometimes form a functional cell aggregate, in which cells strongly interact with each other by exchanging chemicals, as demonstrated with quorum sensing \cite{Sociomicrobiology}. 
Such a ``multi-cellular aggregate'' is often observed to form under a limited resource condition, which may indicate that formation of an aggregate is a universal strategy for a unicellular organism to survive in a severe environment and for cells to grow collectively and cooperatively. 
Interestingly, mutualistic behaviors, cell differentiation, and division of labor 
are ubiquitously observed in such aggregates with isoclonal cells 
\cite{Shapiro2,MCO,Jefferson,pylori,MonospeciesBiofilm,biofilm} 
as well as with heterogeneous cells 
(e.g., bacterial ecosystem) \cite{Shapiro2,MCO,Jefferson,Interspecies,BacterialCoaggregation,PhysioHeterogeneity}. 
This raises the questions of how aggregates of identical cells achieve division of labor for cooperative growth, and what are the necessary conditions? These questions are important to be addressed in order to understand the formation of multicellular aggregates, including the formation of biofilms, 
which has attracted much attention recently \cite{Shapiro2,MCO,Jefferson}.\\ 
\ \ From this point of view, there are at least three characteristics required to 
achieve stably growing aggregates with division of labor.\\ 
(i) Cell differentiation: starting from a single cell, cell states become differentiated as their numbers increase and they coexist and grow.\\
(ii) Cooperative growth: these different cell types mutually cooperate for their growth, possibly by division of labor, 
to avoid being driven to extinction by competition with the isolated, undifferentiated cells.\\ 
(iii) Robustness in the distribution of the number of differentiated cells: balance in the population distribution of cell types is maintained so that cells of different types can coexist stably and grow together.\\
\ \ To achieve stable task differentiation, (i) cell differentiation through cell-cell interaction would be necessary, whereas (ii) cooperative growth is also required, since otherwise community formation through cell-cell communication would not be advantageous (or might even be deleterious) compared with the case of isolated cells without any interaction. 
However, simply achieving this cooperative growth does not necessarily imply that this state is robust, since if one cell type reproduces faster than any other type, the fastest type would dominate the population and 
the appropriate cell type ratio for division of labor would be easily lost. 
Therefore, (iii) coexistence of diverse cell types is also an important issue to be addressed for the stability and survival of a cell colony. \\
\ \ Indeed, such characteristics have also been studied as a primitive form of multicellularity. In experimental evolution, 
aggregation of unicellular organisms with division of labor has been recently investigated with the use of yeast \cite{ratcliff} and algae \cite{ratcliff2}. 
With respect to theoretical approaches, a related issue of the survival of an aggregate of cells has been conventionally discussed in multi-level evolution theory, by introducing a fitness parameter at the cellular and multi-cellular levels, and investigating how these two fitness values are aligned \cite{michod,Roze,Nowak,Willensdorfer,Gavrilets,michod2,Peer}. 
In most of the previous studies based on the prescribed fitness, however, the existence of differentiated cell types is presumed, and thus the capacity of cell differentiation and the fitness alignment are separated \cite{michod,Roze,Nowak,Willensdorfer,Gavrilets,michod2,Peer,Nowak2,Mahadevan}. 
Related criticisms of these previous approaches are discussed in \cite{Doebeli,Robot}. Specifically in \cite{Doebeli}, intracellular dynamics is introduced as the optimization of resource allocation to different tasks under a given artificial fitness function, and it is shown that division of labor emerges when it increases the fitness. 
However, in nature, in general, cell differentiation does not result from optimization of a given fitness but rather results from intra-cellular metabolic reaction dynamics, and thus the division of labor is not guaranteed even when it increases fitness. 
The fitness, i.e., the rate of cellular growth, is also obtained through the reaction dynamics. 
Therefore, it will be important to take cell differentiation and growth rate into account simultaneously, as a result of intra-cellular reaction dynamics, where the growth rate of each cell type is not predetermined, but rather changes according to the cellular states. Furthermore, this growth state also depends on the states of surrounding cells, which may alter the abundances of available resources and the strength of cell-cell interactions. 
To consider these issues that have not been addressed in the previous studies, we here 
present 
a dynamical-systems model of cells with intracellular reactions, cell-cell interactions, and uptake of resources, by which the fitness is determined as the cellular growth rate, rather than being prescribed in advance.\\
\ \ In fact, such models of interacting and growing cells with 
intracellular reaction dynamics 
have been introduced previously, where the concept of isologous diversification \cite{KY1999} has been proposed, to address differentiation from a single cell type (property (i)). 
A previous mathematical model \cite{FK2000} demonstrated that an ensemble of cells sharing a common genotype could achieve niche differentiation through cell differentiation, and thereby relax the strength of resource competition. 
Although this indirect cooperation through niche differentiation would be 
beneficial for cell aggregates, we here address 
cooperative growth in a stronger sense,
where differentiated cells help each other so that interacting cells in an aggregate grow faster than the isolated undifferentiated cells (unicellular organisms) under the condition of limited resources (property (ii)). 
For this purpose, we here consider an environment in which only a single resource exists, and in such situation, property (ii) is considered as the property of the cell ensemble to help each other 
achieve 
a higher growth than the isolated cells, rather than specializing to each resource.\\ 
\ \ In the present paper, by using a simple model of cells that contain diverse components and interact with each other through the exchange of chemicals, we address the question of whether the above three characteristic behaviors are a necessary outcome of an ensemble of interacting cells. 
Specifically, we show that a cell ensemble under strong cell-cell interactions with limited resources fulfills cell differentiation, cooperative growth, and robustness in the cell type distribution. 
\section{
{\large Model}
}
We consider a mathematical model proposed in \cite{FK2000,KY1999,FK1998,FK2001}, which describes a simple, primitive cell that consists of $k$ chemical components \{$X_0, \dots, X_{k-1}$\}. As illustrated in Fig 1, we assume that $n$ cells globally interact with each other in a well-mixed medium, and each of them grows by uptake of the nutrient chemical $X_0$, which is supplied into the medium from the external environment. The internal state of each cell is characterized by a set of variables $(x^{(m)}_0, \dots, x^{(m)}_{k-1}, v^{(m)})$, where $x^{(m)}_i$ is the concentration of the $i$-th chemical $X_i$, and $v^{(m)}$ is the volume of the $m$-th cell ($m = 1, \dots, n$). As a simple model, we consider a situation with only catalysts and resources, where these $k$ components are mutually catalyzed for their synthesis, thus forming a catalytic reaction network. A catalytic reaction from a substrate $X_i$ to a product $X_j$ by a catalyst $X_l$, as $X_i + {\alpha}X_l {\to} X_j + {\alpha}X_l$, occurs at a rate ${\epsilon}{x^{(m)}_{i}}{x^{(m)}_{l}}^{\alpha}$, where $\alpha$ refers to the order of the catalytic reaction and is mostly set as $\alpha = 2$. Here, $\epsilon$ is the rate constant for this reaction, and, for simplicity, all the rate constants are equally fixed at $\epsilon = 1$. 
The parameters and variables in this model are listed in Table \ref{tab:parameters}.\\ 
\begin{figure}[t]
\includegraphics[width = 10 cm]{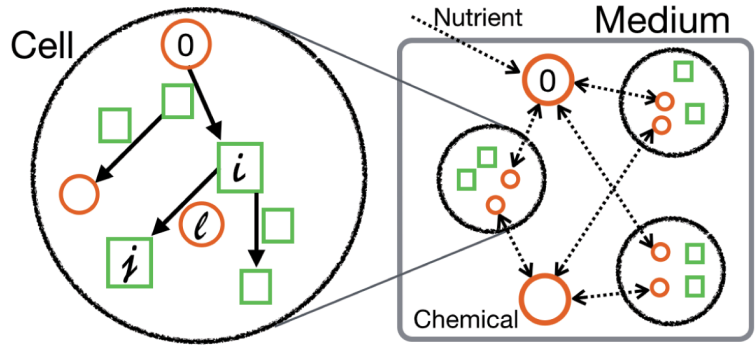}%
\caption{
Schematic illustration of the $N$-cell model. Each cell has a common catalytic network, while the nutrient $X_0$ transported from the medium is transformed to catalytic components for cellular growth. 
The nutrient $X_0$ is supplied to the medium from the exterior. 
The reaction from $X_i$ to $X_j$ is catalyzed by another component, $X_l$. Some components (orange circles) are diffusible and exchanged via the medium, while others (green squares) are not. In the medium, $n$ cells coexist and interact with each other ($1{\leq}n{\leq}N$).
\label{fig:model}}
\end{figure}
\begin{table}[b]
	 \renewcommand{\thetable}{1}
    \caption{The parameters and variables of the N-cell model.}
  \begin{center}
    \begin{tabular}{l|c|c} \hline
      Symbol & Value & Interpretation \\ \hline
      $N$ & $100, 1$ & The maximum number of cells coexisting in a medium\\
      $k$ & $20$ & The number of chemicals\\
      $\alpha$ & $2$ & The order of catalytic reactions \\ 
      $\epsilon$ & $1$ & The rate constant for catalytic reactions \\
      $C$ & $0.05, 0.15$ & The concentration of the nutrient $X_0$ in the medium's exterior \\ 
      $V_{med}$ & $100$ & The volume of the medium \\ 
      $D$ & $0.1, 1$ & The diffusion coefficient between cells and the medium \\ 
      $D_{med}$ & $0.1, 1$ & The diffusion coefficient between the medium and its exterior \\ 
      $P(i,j,l)$ & $0, 1$ & Whether or not the reaction $X_i + {\alpha}X_l {\to} X_j + {\alpha}X_l$ exists \\
      $x^{(m)}_{i}$ & variable & The concentration of the $i$-th chemical $X_i$ in the $m$-th cell\\
      $x^{(med)}_{i}$ & variable & The concentration of the $i$-th chemical $X_i$ in the medium\\ 
      $v^{(m)}$ & variable & The volume of the $m$-th cell\\
       $\mu^{(m)}$ &  variable & The growth rate of the $m$-th cell \color{black} \\ \hline
    \end{tabular}
    \label{tab:parameters}
  \end{center}
\end{table}
\ \ Cell states change through intracellular biochemical reaction dynamics and the in- and outflow of chemicals, leading to cell-cell interactions via the medium. The medium's state is characterized by concentrations $(x^{(med)}_0, \dots, x^{(med)}_{k-1})$, and a constant volume $V_{med}$. 
Then, the dynamics of the concentration of $X_i$ in the $m$-th cell are represented as: 
\begin{eqnarray}
\frac{dx^{(m)}_{i}}{dt} =\sum_{j,l=0}^{k-1} \epsilon{P(j,i,l)}{x^{(m)}_{j}}{x^{(m)}_{l}}^{\alpha} - \sum_{j,l=0}^{k-1} \epsilon{P(i,j,l)}{x^{(m)}_{i}}{x^{(m)}_l}^{\alpha} + D\sigma_{i}({x^{(med)}_{i}} - {x^{(m)}_{i}})  -  {x^{(m)}_{i}}\mu^{(m)},
\end{eqnarray}
where $P(i,j,l)$ takes the value 1 if there is a reaction $X_i + {\alpha}X_l {\to} X_j + {\alpha}X_l$, and is 0 otherwise. In Eq. (1), the third term describes the influx of $X_i$ from the medium, and the fourth term gives the dilution owing to the volume growth of the cell, and $\mu^{(m)}$ denotes the cellular growth rate. 
Here, only a subset of chemical species is diffusible across the cell membranes with the rate of diffusion $D$. 
$X_i$ is transported from the medium into the $m$-th cell at a rate $D\sigma_{i}({x^{(med)}_{i}}-{x^{(m)}_{i}})$, where $\sigma_{i}$ is 1 if $X_i$ is diffusible, and is 0 otherwise. 
Therefore, the $m$-th cell grows in volume according to the rate $\mu^{(m)}{\equiv}\sum_{i=0}^{k-1}{D\sigma_{i}({x^{(med)}_{i}}-{x^{(m)}_{i}})}$ by assuming that the cellular volume is in proportion to the total amount of chemicals. The volume dynamics are given by $dv^{(m)}/dt=\mu^{(m)}v^{(m)}$. As the abundances of chemicals are conserved by the intracellular reactions, with this form of volume growth, $\sum_{i=0}^{k-1}x_i^{(m)}=1$ is time-invariant \cite{FK1998}. 
\ \ The nutrient chemical $X_0$, 
which is necessary for cellular growth, 
is supplied into the medium from the external environment according to the rate ${D_{med}}(C - x^{(med)}_0)$, where ${D_{med}}$ denotes the diffusion coefficient of the nutrient across the medium's boundary, whereas $C$ is the constant external concentration of the nutrient $X_0$ 
(for simplicity, the flow of the other diffusible chemicals to the outside of the medium is not included, although its inclusion does not alter the result below as long as their $D_{med}$ values are not large).\\ 
\ \ Therefore, the temporal change of $x^{(med)}_{i}$ is given by
\begin{eqnarray}
  \frac{d{x^{(med)}_{i}}}{dt}= D_{med}&\sigma^{\prime}_{0}&(C - {x^{(med)}_{i}}) - \sum_{m=1}^{n}\frac{D\sigma_{i}({x^{(med)}_{i}}-{x^{(m)}_{i}}){v^{(m)}}}{V_{med}},
\end{eqnarray}
where $\sigma^{\prime}_{0}$ takes unity only if $i=0$, i.e., if $X_i$ is the nutrient. For simplicity, $D_{med}$ was set as $D_{med}=D$, though the results reported here do not greatly depend on the value of $D_{med}$.\\
\ \ According to these processes, each cell grows 
by converting nutrient chemicals into non-diffusible chemicals and storing them within the cell 
until its volume doubles, and then divides into two cells with almost the same chemical compositions. Here, the catalytic network in daughter cells is identical to that in their mother cell. As the initial condition, only a single cell exists with a randomly determined chemical composition. In addition, we set the carrying capacity of a medium $N$, which is an upper limit to the number of cells that can coexist in the medium. When the cell number exceeds its upper limit $N$ due to cell division, the surplus cells are randomly eliminated. Hereafter, this model is referred to as the $N$-cell model.
\section{
{\large Results}
}
\subsection{Cell differentiation in the N-cell model: brief summary}
We simulated the $N$-cell model over hundreds of randomly generated reaction networks. Each catalytic network is generated in the following manner. Each chemical is set to be diffusible with probability $q=0.15$ and has $\rho=4$ outward reaction paths to other chemicals; i.e., each chemical works as a substrate in $\rho$ reactions. Each reaction $X_i + {\alpha}X_l {\to} X_j + {\alpha}X_l$ ($i \neq j$, and $X_j$ and $X_l$ are not nutrients) is randomly determined so that $j{\neq}l$ is fulfilled. We did not allow for autocatalytic reactions ($j=l$) as they are rare in nature. However, the described results were also obtained without these restrictions.\\ 
\ \ We are particularly interested in if and how the cells differentiate, and whether the growth rate would increase as a result of differentiation. For this purpose, cell differentiation is defined as the emergence of cells with different chemical compositions within the population that share an identical catalytic network. 
For the case where the concentrations asynchronously oscillate in time, we evaluated whether cells have different compositions even after taking the temporal average over a sufficiently longer time scale than the oscillation period. To evaluate the growth enhancement, we compared two different situations, ``interacting" ($N=100$) and ``isolated" ($N=1$) cases, and then we computed $R_{\mu}$, the ratio of the growth rate of interacting cells to that of isolated cells. Then the growth enhancement is defined as $R_{\mu}>1$.\\
\ \ The behavior of the $N$-cell model is classified into four categories. 
In category (a), interacting cells differentiate into two or more types and grow faster than isolated cells, i.e., $R_{\mu}>1$ (Fig 2; see also Figure A in S1 Text)
In category (b), interacting cells differentiate but their growth is slower than that of isolated cells ($R_{\mu}<1$); 
in this category, as far as we have examined, cells of a certain type gain chemicals diffused from another type, which are used as catalysts for conversion to non-diffusible chemicals. The latter cell type has a composition similar to that of the isolated cell, and its growth is decreased by this cell-cell interaction (see Figure B in S1 Text). 
Hence, the former cell type is considered to exploit the latter as it receives the unidirectional chemical inflow. 
In category (c), cells do not differentiate with respect to the average composition, but chemical concentrations asynchronously oscillate in time. In category (d), the behavior of each cell is identical, regardless of the presence or absence of cell-cell interactions, and therefore $R_{\mu}=1$.\\
\begin{figure}[t]
\includegraphics[width = 13 cm]{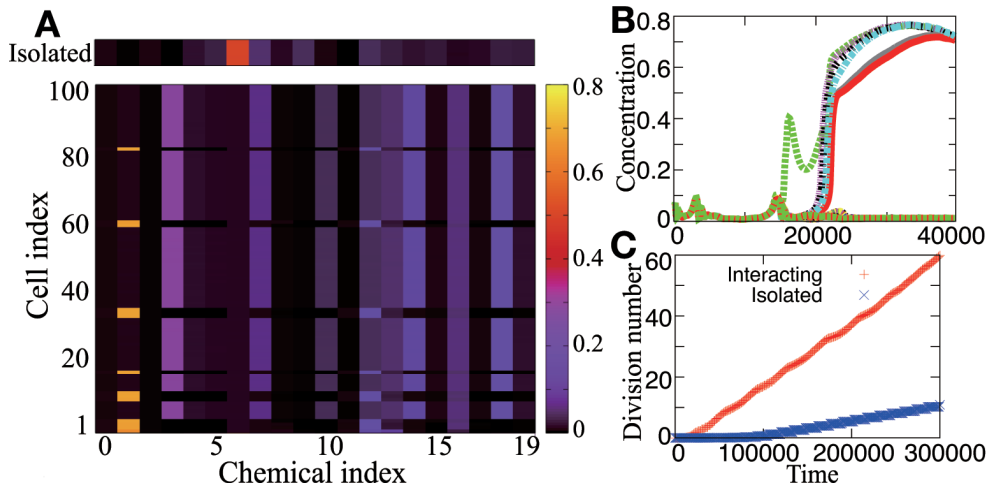}%
\caption{
Typical behavior of the $N$-cell model in category (a). 
(A) Chemical compositions of $N$ cells surviving at time $t =5{\times}10^5$. The concentration $x_i^{(m)}$ is plotted with a color code, with the vertical axis indicating the cell index $m$, and the horizontal axis indicating the chemical index $i$, while the top band designates the composition of an ``isolated" cell. Cells differentiate into two types with distinct compositions. (B) The time series of $x_1^{(m)}$ in interacting $N$ cells surviving at time $t =4{\times}10^4$, overlaid for all cells shown as different colors. 
Each line represents the concentration of the chemical $X_1$ in a different cell, plotted with different colors over 100 cells.
(C) The time series of the number of cell divisions per cell for interacting (red) and isolated (blue) cells. 
Interacting cells grow faster than isolated cells; i.e., $R_{\mu}>1$.
\label{fig:model}}
\end{figure}
\ \ Here, we are mainly concerned with category (a), as this case enables both cell differentiation and cooperative growth. We found four common properties in this category. (1) A state with homogeneity among cells becomes unstable as the cell number increases, and is replaced by two (or more) distinct cellular states. (2) In differentiated cells, the compositions are concentrated for only a few chemicals, whereas the concentrations of the other chemicals are nearly zero; i.e., each cell type uses only a sub-network of the total reaction network. 
(3) Different cell types share only a few common components, and each of the other components mostly exists in one cell type. 
(4) The components that predominate in one cell type diffuse to the other cell type, where they function as catalysts, and vice versa. 
Thus, the two cell types help each other to achieve higher cooperative growth.
\subsection{Reduction of the N-cell model}
After examining a number of networks in category (a), we extracted a common core structure in the reaction network topology, designated as networks 1-3 (Figs 3A and 3B; see also Figure C in S1 Text). 
In these networks, cells in the $N$-cell model differentiate into two types, 
type-1 and type-2, 
as exemplified in Fig 3C. In type-1, $x_1$ is high while $x_2$ is close to zero, and in type-2, $x_2$ is high and $x_1$ is close to zero. Accordingly, $X_3$ ($X_4$) can be produced only in the former (latter) type, and the two types of cells complement each other by exchanging $X_3$ and $X_4$. Consequently, the differentiated cells grow faster than the isolated cells (Fig 3D).\\
\begin{figure}[t]
\includegraphics[width = 13 cm]{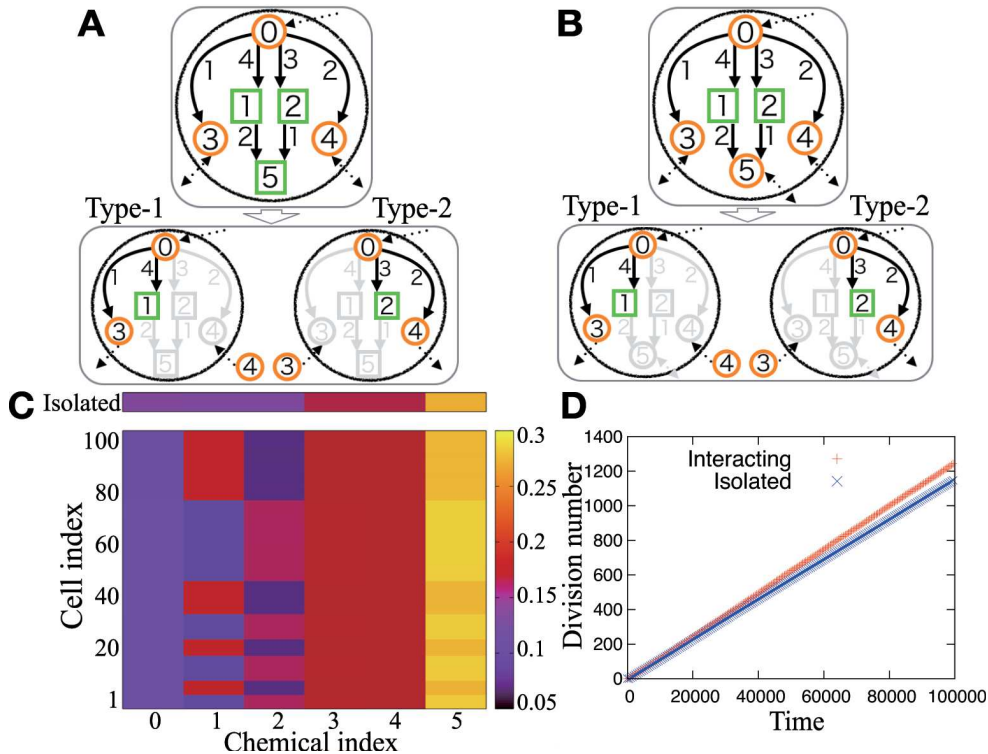}%
\caption{
Behavior of simplified networks. 
(A) Network 1. (B) Network 2. 
In (A) and (B), each number represents the chemical index, with orange-circle and green-square nodes representing diffusible and non-diffusible chemicals, respectively. The difference between networks 1 and 2 lies only in the diffusibility of $X_5$. The dashed arrows denote  the diffusive fluxes of chemicals, and the thick arrows indicate catalytic reactions. The chemical at the arrowtail is transformed to the chemical at the arrowhead, catalyzed by the chemical labeled on the edge; e.g., the left arrow from $X_0$ in (A) denotes the catalytic reaction $X_0 + {\alpha}X_1 {\to} X_3 + {\alpha}X_1$. When the cells differentiate, type-1 (type-2) cells mainly produce the chemicals $X_1$ and $X_3$ ($X_2$ and $X_4$), and receive $X_4$ ($X_3$) from type-2 (type-1) cells, which are illustrated in the lower panel with color. 
(C) and (D) are examples of the behavior of network 1 for $(C,V_{med},D)=(0.15,100,1)$. (C) Chemical concentrations $x_i^{(m)}$ in $N$ cells surviving at time $t = 10^5$ are plotted according to the color code shown in the sidebar, with the vertical axis as the cell index $m$ and the horizontal axis as the chemical index $i$. The top band designates an ``isolated" cell. (D) The time series of the number of cell divisions per cell for interacting (red) and isolated (blue) cells.}
\end{figure}
\ \ To analyze the mechanism of this cooperative differentiation, we reduced the $N$-cell dynamics to two effective groups of cells represented by $(x^{(i)}_0, \dots, x^{(i)}_{k-1}, v^{(i)})$, where $v^{(i)}$ denotes the total volume of each cell group ($i=1,2$). Considering that the total cell number is sustained at its maximum $N$, the total cellular volume is also bounded. Therefore, $v=v^{(1)}+v^{(2)}$ is regarded as a constant in the reduced version of interacting cells, termed the reduced-2cell (r2cell) model. This model obeys 
Eqs. (1)-(2) with $\mu^{(m)}{\equiv}\sum_{i=0}^{k-1}{D\sigma_{i}({x^{(med)}_{i}}-{x^{(m)}_{i}})}$, 
and the effect of random cell elimination associated with cell division is implicitly incorporated into dilution due to volume growth. 
Besides, by considering the symmetry in networks 1-3, we can assume $v^{(1)}=v^{(2)}=v/2$ for symmetric differentiation with the same number of cells of the two types, while the case with $v^{(1)}{\neq}v^{(2)}$ will be discussed later.\\
\ \ Likewise, we also consider the reduced-1cell (r1cell) model corresponding to the ``isolated" case of the $N$-cell model, by ignoring cell division and assuming that the cellular volume is constant at $v^{(iso)}=v$.
\subsection{Cell differentiation} 
The behavior of the r1cell and r2cell models (i.e., isolated and interacting cells) can be classified into several phases, depending on parameters $(C, V, D)$, where $V{\equiv}V_{med}/v$ is the volume ratio between the cells and the medium.\\ 
\ \ The phase diagram with network 1 for $D=1$ is shown in Fig 4A, and Figure E in S1Text shows phase diagrams of networks 1-3 for various $D$ values. 
The blue area in Fig 4A represents phase (I), in which the cells cannot differentiate, and always reach a single fixed point attractor in both the r1cell and r2cell models. In phase (II), differentiation into two fixed points occurs in the r2cell model from a stable fixed point in the r1cell model, as shown in Fig 4B. In phase (III), the r1cell model exhibits oscillation, while two cells in the r2cell model reach two distinct fixed points (Fig 4C). 
In terms of dynamical systems theory, this loss of oscillation is referred to as oscillation death \cite{oscillator death, oscillation death}. In phase (IV), both ``oscillation-death" differentiation and synchronous oscillation (i.e., non-differentiation) can occur depending on the initial condition, whereas the r1cell model always exhibits oscillation.\\
\begin{figure}[t]
\includegraphics[width = 11.5 cm]{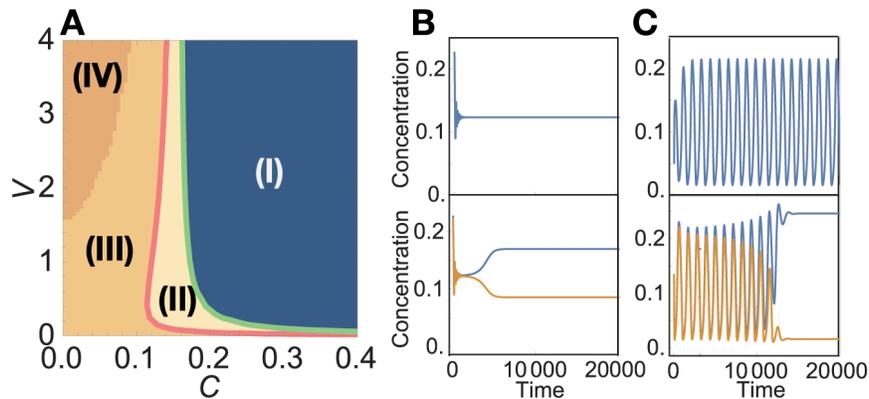}%
\caption{
The behavior of the r1cell and r2cell models with network 1. 
(A) A phase diagram ($D=1$). The blue area designates phase (I) in which cells cannot exhibit differentiation, and reach a single fixed point in both the r1cell and r2cell models. In phases (II) and (III), cells always differentiate. In phase (II), shown in cream color, cells exhibit pitch-fork bifurcation from a homogeneous state. In phase  (III), ``oscillation-death" occurs in the dynamics of the chemical compositions of the two cells. In phase (IV), depicted by orange, if the initial difference between the compositions of the two cells is large enough, ``oscillation-death" occurs; otherwise, the r2cell model exhibits non-differentiated, synchronized oscillation. (B) The time series of concentrations of $X_1$ in phase (II) for $(C,V,D)=(0.27,0.1,1)$. (C) The time series in phase (III) for $(C,V,D)=(0.1,0.8,0.1)$. For both (B) and (C), the time series for the r1cell (r2cell) model is displayed on the upper (lower) column.}
\end{figure}
\ \ Thus, differentiation occurs in phases (II)-(IV) (i.e., at the left of the green line in Fig 4A), 
while stable differentiation without falling into synchronized oscillation is achieved 
only in phases (II)-(III), i.e., with small $C$ and $V$ values, representing a limited resource and strong cell-cell interaction condition. 
In Fig 4A, phases (II)-(III) are divided by the red line, and the red and green lines are determined according to linear stability analysis (see S1 Text for details). \\
\ \ With respect to the network structure, the catalytic reactions $X_1+{\alpha}X_2{\to}X_5+{\alpha}X_2$ and $X_2+{\alpha}X_1{\to}X_5+{\alpha}X_1$ function as two mutually repressive reactions, i.e., forming a double-negative feedback loop. 
Further, the product $X_5$ consumes $X_1$ and $X_2$, and is maintained within the cell, which enhances the dilution of $X_1$ and $X_2$. Thus, each of these reactions works as a composite negative feedback loop, leading to instability of the homogeneous cell state. Since nonlinearity is a necessary condition for multi-stability, a high order of catalytic reactions $\alpha$ tends to facilitate cell differentiation.
\subsection{Cooperative growth by differentiation}
Fig 5A shows the dependence of $R_{\mu}$ on parameters in network 1, exemplifying that differentiation increases the growth rate. 
Surprisingly, this differentiation-induced growth enhancement was always observed for any set of parameters in networks 1-3 (network 3 is shown in Figure C in S1 Text).\\
\begin{figure}[b]
\includegraphics[width = 11 cm]{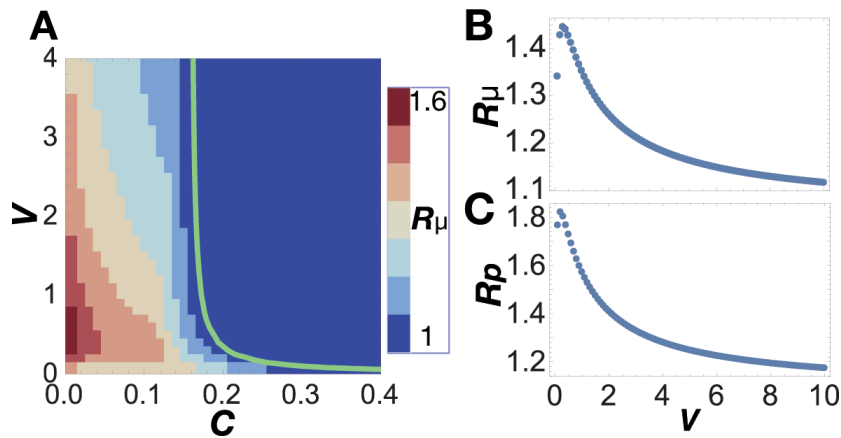}%
\caption{
Growth enhancement with network 1 ($D=1$). 
(A) Dependence of $R_{\mu}$ on parameters $C$ and $V$. Cells can differentiate given the parameters to the left of the green line. (B) Dependence of $R_{\mu}$ on $V$ ($C=0.1$). $R_{\mu}$ is the ratio of the growth rate of differentiated cells to that of isolated cells. (C) Dependence of $R_p$ on $V$ ($C=0.1$). $R_p$ is the degree of increase in the production of $X_3$-$X_4$ conferred by differentiation.}
\end{figure}
\ \ We next sought to determine the mechanism contributing to the faster growth of differentiated cells. An intuitive explanation is as follows. On one hand, an isolated cell must contain all chemical components required for self-reproduction (e.g., $X_0$-$X_5$ in the upper panel of Fig 3A), leading to lower concentrations of each chemical on average. On the other hand, differentiated cells can achieve division of labor; each type of differentiated cell exclusively produces a portion of the required chemical species, and cells exchange these chemicals with each other. Since catalytic reactions occur only in a sub-network of the original network (e.g., a network in the lower panel of Fig 3A), the chemicals are concentrated on fewer components, which increases the efficiency of chemical reactions and promotes cellular growth.\\ 
\ \ This suggests that stronger cell-cell interactions support higher growth. Indeed, Fig 5B shows that a smaller $V$, i.e., stronger cell-cell interaction, causes larger $R_{\mu}$. A smaller $V$ also increases $R_p$, the ratio of the total production of $X_3$-$X_4$ in the r2cell model to that in the r1cell model (Fig 5C);
that is, the production of exchanged chemicals is enhanced. 
To conclude, stronger cell-cell interactions reinforce the division of labor, whereby differentiated cells can grow more efficiently.\\ 
\ \ The rate of growth enhancement through cell differentiation $R_{\mu}$ can be roughly estimated by recalling that the growth rate of a cell is given by the average influx of the nutrient chemical it receives. We compared the growth rate of an isolated cell $\mu^{(iso)}$ to that of a differentiated cell $\mu^{(dif)}$ by assuming that the concentration of each chemical species is equally distributed, except for the nutrient chemical. 
Considering that an isolated cell 
has a catalytic network with $k$ chemical components 
and $q$ reaction paths from the nutrient $X_0$, each concentration of $X_1$-$X_{k-1}$ is calculated as $x^{(iso)}=(1-x^{(iso)}_0)/(k-1)$. Hence, for the steady state, the growth rate is estimated by $\mu^{(iso)}=q{x^{(iso)}_0}{x^{(iso)}}^{\alpha}/(1+x^{(iso)}_0)$. On the other hand, the sub-network in a differentiated cell is considered to have $k^{\prime}$ chemicals and $q^{\prime}$ reaction paths from the nutrient ($k^{\prime}<k$, $q^{\prime}<q$). Then, each chemical concentration and the growth rate are given by $x^{(dif)}=(1-x^{(dif)}_0)/(k^{\prime}-1)$ and $\mu^{(dif)}=q^{\prime}{x^{(dif)}_0}{x^{(dif)}}^{\alpha}/(1+x^{(dif)}_0)$, respectively.\\
\ \ Here, we also assume that $x^{(iso)}_0{\simeq}x^{(dif)}_0$, because these concentrations mostly depend on the supplied nutrient concentration $C$ rather than on the internal dynamics of individual cells. From these assumptions, the growth ratio $R_{\mu}{\equiv}\mu^{(dif)}/\mu^{(iso)}$ is calculated as $R_{\mu}=(q^{\prime}/q)[(k-1)/(k^{\prime}-1)]^{\alpha}$. For example, with network 1 or 2 (Figs 3A and 3B), 
$k=6, q=4, k^{\prime}=4, q^{\prime}=2$, and thus $R_{\mu}=(1/2){(5/3)}^{\alpha}$, which is greater than unity, at least when $\alpha{\geq}2$. Although this estimate is not strictly accurate, it nevertheless demonstrates how cell differentiation can enhance cellular growth, which is facilitated by greater $\alpha$. Even when the chemical concentrations were non-uniform, division of labor could accelerate growth when $\alpha$ was sufficiently large.
\subsection{Robustness in the number distribution of differentiated cells}
The cells in our models achieved (i) cell differentiation and (ii) cooperative growth. 
However, if one cell type grows faster than the other type, 
the cooperation between the differentiated cells collapses. Thus, the third condition is necessary: the growth rate of each cell type needs to be in conformity, through mutual regulation by cell-cell interactions.\\ 
Thus far, we have considered the case with equal populations of the two cell types by imposing the condition $v^{(1)}=v^{(2)}$. Here, we examine the case with $v^{(1)}{\neq}v^{(2)}$ for fixed $v^{(1)}$ and $v^{(2)}$, to evaluate whether the increases in cell volume (or number) are identical between the two types to meet the requirement (iii). 
Therefore, Figs 6A and 6B show plots of the growth rate versus $r^{(1)}$, where $r^{(i)}{\equiv}v^{(i)}/(v^{(1)}+v^{(2)})$ is the volume proportion between type-1 and type-2 cells. Now, let us denote the dependence of $\mu^{(1)}$ on $r^{(1)}$ by a function $F(r^{(1)})$. 
Then, the growth rate of the type-2 cell $\mu^{(2)}$ is given by $G(r^{(2)})=G(1-r^{(1)})$, which is equal to $F(1-r^{(1)})$ due to symmetry in the catalytic network.\\ 
\begin{figure}[t]
\includegraphics[width = 12 cm]{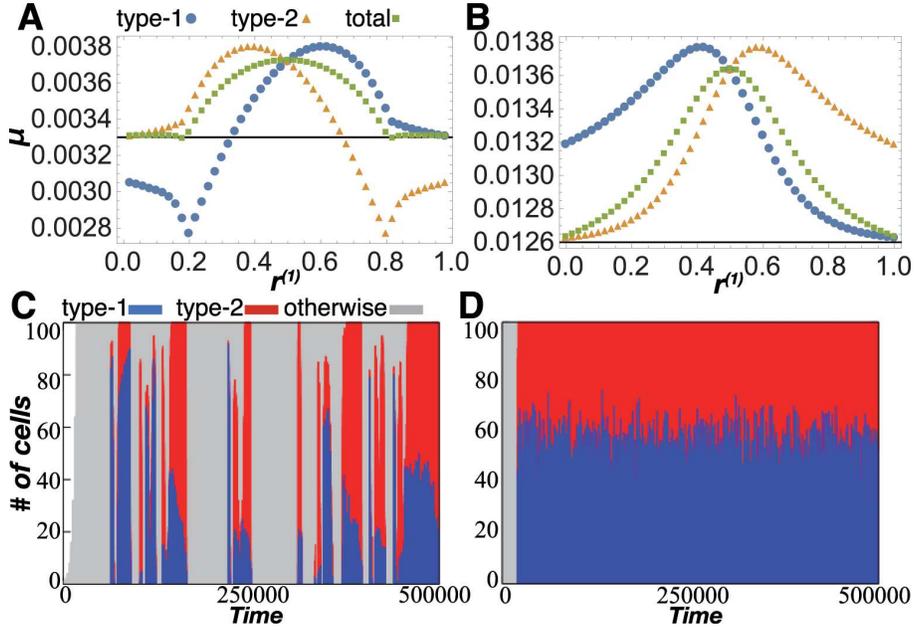}%
\caption{
The stability of collective growth. 
(A) Network 1 in the r2cell model for $(C,V,D)=(0.1,1,0.1)$, representing unstable collective growth. (B) Network 2 in the r2cell model for $(C,V,D)=(0.15,3,1)$, representing balanced collective growth. In (A) and (B), the horizontal axis denotes the volume ratio of the two cells, $r^{(1)}{\equiv}v^{(1)}/(v^{(1)}+v^{(2)})$, and each black line displays the growth rate in the r1cell model $\mu^{(iso)}$. Blue circles and orange triangles denote the growth rate of type-1 $\mu^{(1)}$  and type-2 $\mu^{(2)}$, respectively. Green rectangles indicate the total growth rate of the cell aggregate $\mu^{(tot)}{\equiv}{r^{(1)}}{\mu^{(1)}}+r^{(2)}{\mu^{(2)}}$. (C) and (D) show the cell number distribution in the $N$-cell model ($N=100$). Blue, red, and gray bars designate the numbers of type-$1$, type-$2$, and non-differentiated cells, respectively. (C) Case for network 1 with $(C,V_{med},D)=(0.01,100,0.1)$. (D) Case for network 2 with $(C,V_{med},D)=(0.3,50,0.1)$.}
\end{figure}
\ \ Since differentiated cells help each other, balanced growth would be expected; if the volume or relative number of one cell type is larger than the other, a larger (smaller) amount of chemicals would be supplied from the majority (minority) type to the minority (majority) type, so that the growth rate of the minority type is enhanced compared to that of the other type. This is the case for network 2, where $F(r^{(1)})<F(1-r^{(1)})$ for $r^{(1)}>1/2$, and the difference in volume (or number) decreases over time, leading to a balanced cell distribution (Fig 6B).\\ 
\ \ Nonetheless, this is not always the case. Fig 6A shows that the growth rate of the majority type is larger than that of the minority type in network 1; that is, $F(r^{(1)})>F(1-r^{(1)})$ for $r^{(1)}>1/2$. Accordingly, the difference in volume increases over time, and thus the two different cell types cannot stably coexist. This instability of collective growth differs from a scenario of parasitic behavior, because $\mu^{(tot)}{\equiv}{r^{(1)}}{\mu^{(1)}}+r^{(2)}{\mu^{(2)}}$ is higher than $\mu^{(iso)}$ for almost the entire range of $r^{(1)}$.\\ 
\ \ The condition for the stability is analytically expressed as follows. First, the temporal change of $r^{(1)}$ is represented by 
$dr^{(1)}/dt =r^{(1)}(1-r^{(1)})[F(r^{(1)})-F(1-r^{(1)})]$, 
which has a trivial fixed point solution $r^{(1)}=1/2$. This fixed point, where the two cell types coexist, is unstable if $F^{\prime}(1/2)>0$, and is stable if $F^{\prime}(1/2)<0$. To estimate $F^{\prime}(1/2)$, recall that $dx^{(m)}_i/dt=0$ is fulfilled in a cell with steady growth, and $D_{med}V{\gg}D$. Then, from the definition of the growth rate $\mu$, we get 
\begin{eqnarray}
  \frac{F^{\prime}(1/2)}{D} {\simeq} - \frac{{\partial}x^{(1)}_0}{{\partial}r^{(1)}}-\frac{1}{2}\sum_{i=1}^{k-1}{\sigma_{i}} \left. \frac{{\partial}(x^{(1)}_i-x^{(2)}_i)}{{\partial}r^{(1)}}\right|_{r^{(1)}=\frac{1}{2}}.
\end{eqnarray}
Since the first term is always negative as described in S1 Text, Eq. (3) shows that if the difference in exchanged chemicals between the majority and minority cells [$\sum_{i=1}^{k-1}{\sigma_{i}}(x^{(1)}_i-x^{(2)}_i)$] increases in proportion to the increase in volume ratio of the majority type, then $F^{\prime}(1/2)$ is negative and thus the collective growth is balanced.\\
\ \ Now, let us consider how the difference in volume alters the states of two interacting cells in networks 1 and 2. When $r^{(1)}>1/2$, 
the population ratio of type-1 cells is increased, and the amount of $X_4$ supplied from the minority type-2 
cells is not sufficient for the majority type-1 cells to maintain their differentiated chemical composition. In contrast, the minority type-2 cells receive sufficient amounts of $X_3$ from the majority type-1 cells, and maintain their differentiated composition. Consequently, the chemical composition of the type-1 cells approaches that of the isolated case, which contains $X_1$ and $X_2$ equally. This indicates that the majority of type-1 cell produces more $X_5$ than the minority of type-2 cell does, and thus $\partial{(x_5^{(1)}-x_5^{(2)})}/ \partial r^{(1)}> 0$ holds around $r^{(1)}=1/2$. Therefore, the diffusibility of $X_5$ contributes to the stability of network 2, and the non-diffusibility of $X_5$ contributes to the instability of network 1. Intuitively, this mechanism can also be explained as follows: with network 1, the majority cell type can produce a greater fraction of a non-diffusible chemical $X_5$ for itself and a smaller fraction of a diffusible chemical $X_3$ or $X_4$ for the other cell type, and thus the majority cell type grows faster than the minority cell type.\\ 
\ \ The stability and instability of collective growth are also observed in the original $N$-cell model. With an ``unstable" network 1, the $N$-cell model repeats the following dynamic behavior, as shown in Fig 6C: the medium is dominated by cells of one type, and cells of the minority type become extinct. Then, their differentiated compositions cannot be maintained with a single cell type, leading to de-differentiation. Thus, the coexistence of differentiated cells is temporally regained.\\
\ \ In contrast, in a ``stable" network 2, the two differentiated cell types stably coexist and their growth is balanced (Fig 6D), in which a perturbation to increase the population of one cell type leads to a decrease in the growth rate of that type. 
\section{
{\large Discussion}
}
In this paper, we have shown that an aggregate of identical cells achieves metabolic division of labor, with strong cell-cell interactions under limited resources. 
We have revealed how (i) cell differentiation, (ii) growth enhancement, and (iii) robustness in a cell population can be simultaneously self-organized without assuming the ability of differentiation \textit{a priori}, in a simple system consisting only of intracellular reaction dynamics and cell-cell interactions through chemical diffusion.\\ 
\ \ First, cells sharing a common genotype (i.e., an identical reaction network and identical parameters for reaction and diffusion) differentiate into several types with different chemical compositions 
as a result of the instability of a homogeneous cell state induced by cell-cell interactions. This differentiation is facilitated under a condition of limited resources and strong cell-cell interactions, given a high order of catalytic reactions. This dynamical-systems mechanism has also been proposed in previous studies using models of metabolic networks \cite{FK2000} and gene regulation networks \cite{GK2013}.\\
\ \ Second, the differentiated cells can achieve cooperative division of labor. 
The explanation of the division of labor given in the Results section can be simply sketched as follows. Let us consider two reactions in the r2cell model, $X_0 + {\alpha}X_1 {\to} X_3 + {\alpha}X_1$ and $X_0 + {\alpha}X_2 {\to} X_4 + {\alpha}X_2$, with $x_1+x_2=c$. 
If $x_1 = x_2 = c/2$, the total production rate of $X_3$ and $X_4$ is $2x_0(c/2)^{\alpha}$. In contrast, if the concentrations are biased either to $X_1$ or $X_2$ due to differentiation, $x_1 \sim c$ (or $x_2 \sim c$), the total production rate of $X_3$ and $X_4$ per cell is $x_0c^{\alpha}$, which is $2^{{\alpha} -1}$ times greater than that of the previous isolated, generalist cell. Thus the higher the order of the reaction $\alpha$, the greater the advantage of division of labor. A more precise argument is given in the Results (ii). 
This growth enhancement due to division of labor is clearly distinguishable from a scenario of relaxation of the competition for resources through niche differentiation reported previously \cite{FK2000}, in which the growth rate is not increased relative to that of an isolated cell. 
In our model, the growth rate can be enhanced by concentrating chemicals on one of the modules in the network, while the other module is necessary for catalyzing the reaction. 
A related mechanism for division of labor was proposed by Michod and colleagues (see also \cite{Peer,Doebeli,Wagner,DirtyWork} for discussion on the trade-off for division of labor). In the framework of Michod et al., the convexity of the trade-off function is important for division of labor, and the condition of $(q^{\prime}/q)[(k-1)/(k^{\prime}-1)]^{\alpha}>1$ in our model may be related to the convexity of the trade-off function.\\ 
\ \ Although the proposed model describes 
the division of chemical production 
among a cell aggregate, the above mechanism can be seen as analogous to the theory for division of labor in economics: the theory of comparative advantage proposed by Ricardo \cite{Ricardo} states that the mutual use of surplus from a different country is more advantageous than producing all necessary resources in a single country, unless the transport cost is too high. 
In this sense, Ricardo's theory parallels the present mechanism, because two cell types help each other by exchanging the products that are necessary to the other cell type. Indeed, our mechanism works best when cell density is high, so that chemicals are easily exchanged without much loss within the medium. 
Note, however, that in Ricardo's theory, trade is assumed to occur between countries that differ in their relative ability of producing multiple goods, in contrast to the current model in which cells share an identical chemical network. With regard to this point, a better comparison would be with Taylorism \cite{Taylor}, which refers to increases in a group's productivity by each member specializing to each task without assuming individuals with different abilities \cite{KY1999}.\\
\ \ Remarkably, this cooperative differentiation is not sufficient to satisfy condition (iii), robustness in the number distribution of differentiated cells. If one cell type begins to dominate the population, production of the chemicals needed by the minority type will increase. Thus, a feedback mechanism to reduce the majority population is expected. 
However, if the fraction of non-diffusible chemicals is increased for the majority cell type, this storage of chemicals within a cell would suppress the supply of chemicals for the other cell types. Consequently, the majority cell type would further increase its population. 
This suggests that to achieve a balanced population state, the mutual transport of necessary chemicals must work efficiently beyond any possible increase in internal reserves. 
From this perspective, it may be interesting to consider a possible economic analogy: reducing internal reserves and sharing a higher degree of wealth will be relevant to the stabilization of groups with division of labor. 
We here stress that this instability of cooperation can emerge only when cells simultaneously achieve differentiation and division of labor. \\
\ \ In theoretical studies for multi-level evolution, a game theory approach has been sometimes adopted to address the evolution and dynamics of conflict between individuals and society. Although our approach differs from game theory, it might be worth discussing our result in light of this perspective. From a game theory perspective, the cellular growth rate is regarded as a measure of fitness or score. Hence, when two cell types stably coexist in network 2, stable Nash equilibrium is achieved at $r^{(1)}=1/2$. In contrast, in network 1, no stable equilibrium exists for $0<r^{(1)}<1$, and only unstable Nash equilibrium exists at $r^{(1)}=1/2$, 
and thus one type dominates the population. Interestingly, after extinction of one type, re-differentiation of the remaining cells into two types increases the fitness (i.e., growth rate) for both types, as shown in Fig 6A. This dominance of one cell type and re-differentiation are repeated as a result of symbiotic growth and differentiation due to the instability of a homogeneous cell society. 
We expect that such dynamic behavior will be observed in an artificial symbiosis experiment with \textit{Escherichia coli} and diffusible amino acids \cite{HY2011,H2011}.\\
\ \ Considering the difference between networks 1 and 2, the stability and instability of the system can be switched by even a slight change in the diffusibility of a single chemical species. This implies that slight epigenetic changes and transcriptional errors occurring during the lifetime of an organism can lead to instability in the cell distribution, which may relate to the phenomena of metamorphosis during development and carcinogenesis.\\
\ \ Our results demonstrate that an aggregate of simple cells consisting only of catalytic reactions and the diffusive transport of chemicals 
can fulfill differentiation with division of labor, collective growth with symbiotic relationship, and stability. 
Note that these basic characteristics in our model emerge without a fine-tuned mechanism, and are facilitated by the conditions of strong cell-cell interactions, limited resources, and a high order of catalytic reactions.\\ 
\ \ From this point of view, it is interesting to compare the present results to some characteristics of 
multicellular aggregates. 
First, filaments of the cyanobacterium \textit{Anabaena} are known to differentiate, with each cell metabolically specializing in photosynthesis or nitrogen fixation, enabling more efficient growth \cite{Anabaena}. 
Second, such cell differentiation with metabolic division of labor in some cyanobacteris occurs in response to combined nitrogen limitation \cite{Akinete,Heterocyst}. 
 \color{black}
Third, the biofilm of \textit{Bacillus subtilis} exhibits metabolic co-dependence between interior and peripheral cells by chemical oscillation \cite{biofilm}, 
suggesting the relevance of nonlinear dynamics and cell-cell interactions for differentiation. \\ 
\ \ In contrast to the present model of symbiotic growth, however, it has been pointed out that most multicellular aggregates and organisms have achieved division of labor between reproductive and non-reproductive cells throughout evolution \cite{Simpson}. 
Nevertheless, at some developmental stage of multicellular aggregates, symbiotic growth of different cell types may be expected to exist, by achieving differentiation and functional division of labor for biofilm formation \cite{Bacillus subtilis1,Bacillus subtilis2}. Furthermore, in our model, whether or not both types of differentiated cells reproduce strongly depends on the conditions. For example, in network 1, the growth rate is different between the major and minor cell types. Depending on the parameters, there are also cases in which one cell type would cease growing. In addition, we considered the symmetric differentiation case for clarity, but if the reaction rates are different by different chemicals (which are natural), the growth rates of differentiated cell types could generally be different. 
Further, if we assume that cellular growth is determined by a certain chemical (e.g., $X_1$ in networks 1-3) rather than by the total amount of chemicals, after differentiation, one cell type will be reproductive, and the other non-reproductive, while maintaining functional division of labor. \\
\ \ Interestingly, characteristics (i)-(iii) can be part of the requirements for multicellularity. Thus, such characteristics may provide a primitive step to the evolution of multicellular organisms, which has been a topic of much attention from both theorists and experimentalists over the last few decades \cite{MS,Shapiro,0,1,InPress,Ricard,michod,FK2000}. 
In this context, our results are also related to the experimental emergence of multicellularity from unicellular organisms \cite{ratcliff,ratcliff2}. 
However, the three characteristics may not be sufficient for the emergence of multicellular organisms. 
For example, besides the metabolic division of labor, multicellular organisms ubiquitously show germ-soma differentiation and apoptosis. Therefore, determining how the cell aggregates with metabolic division of labor considered here might achieve this universal property of multicellularity remains an important issue to be addressed. 
\section{
{\large Supporting Information}
}
\textbf{S1 Text.} Supplementary information about the description and analysis of the $N$-cell, r1cell, and r2cell models: Figures A and B Typical behavior of categories (a) and (b); Figure C Network 3; Figure D Dependence of the concentration of the nutrient $X_0$ on $r^{(1)}$; Figure E Phase diagrams of the r1cell and r2cell models.

\section{
{\large Acknowledgements}
}
We would like to acknowledge helpful discussions with Chikara Furusawa.

\end{document}